\documentstyle[11pt,aaspp4]{article}
\def\cm{{\rm cm}}

\def\hmpc{\;h^{-1}{\rm Mpc}}

\def\invhmpccc{\;h^{3}{\rm Mpc}^{-3}}
\def\hmpccc{\;h^{-3}{\rm Mpc}^{3}}
\def\hkpc{h^{-1}{\rm kpc}}
\def\kms{{\rm \;km\;s^{-1}}}

\def\lya{Ly$\alpha$\ }

\def\nh1{n_{\rm HI}}

\def\K{{\rm K}}

\def\p1dk{P_{\rm 1D}(k)}
\def\msun{{\,M_\odot}}
\def\nbar{{\overline{n}}}
\newbox\grsign \setbox\grsign=\hbox{$>$} \newdimen\grdimen \grdimen=\ht\grsign
\newbox\simlessbox \newbox\simgreatbox
\setbox\simgreatbox=\hbox{\raise.5ex\hbox{$>$}\llap
     {\lower.5ex\hbox{$\sim$}}}\ht1=\grdimen\dp1=0pt
\setbox\simlessbox=\hbox{\raise.5ex\hbox{$<$}\llap
     {\lower.5ex\hbox{$\sim$}}}\ht2=\grdimen\dp2=0pt
\newcommand{\gta}{\mathrel{\copy\simgreatbox}}
\newcommand{\lta}{\mathrel{\copy\simlessbox}}

\begin{document}
 
\title{The Clustering of High Redshift Galaxies in the Cold Dark 
Matter Scenario}
\author{Neal Katz$^{1,4}$, Lars Hernquist$^{2,5}$, and David H. 
Weinberg$^{3,6}$}
 
\footnotetext[1]{ Department of Physics and Astronomy, 
University of Massachusetts, Amherst, MA, 01003}
\footnotetext[2]{Lick Observatory, University of California, Santa Cruz, 
CA 95064}
\footnotetext[3]{Department of Astronomy, The Ohio State University,
Columbus, OH 43210}
\footnotetext[4]{nsk@kaka.phast.umass.edu}
\footnotetext[5]{lars@helios.ucolick.org}
\footnotetext[6]{dhw@astronomy.ohio-state.edu}
 
\begin{abstract} 

We investigate the clustering of high redshift galaxies in five variants of the
cold dark matter (CDM) scenario, using hydrodynamic cosmological simulations
that resolve the formation of systems with circular velocities
$v_c \geq 100\;\kms$ ($\Omega=1$)
or $v_c \geq 70\;\kms$ ($\Omega=0.4$).  Although the five models differ in
their cosmological parameters and in the shapes and amplitudes of their mass
power spectra, they predict remarkably similar galaxy clustering at $z=2$, 3,
and 4.  The galaxy correlation functions show almost no evolution over this 
redshift range, even though the mass correlation functions grow steadily in
time.  Despite the fairly low circular velocity threshold of the simulations,
the high redshift galaxies are usually highly biased tracers of the underlying
mass distribution; the bias factor evolves with redshift and varies from model
to model.  Predicted correlation lengths for the resolved galaxy population
are $2-3\hmpc$ (comoving) at $z=3$.  More massive galaxies tend to be more 
strongly clustered.  These CDM models have no difficulty in explaining the
strong observed clustering of Lyman-break galaxies, and some may even predict
excessive clustering.  Because the effects of bias obscure differences
in mass clustering, it appears that Lyman-break galaxy clustering will not 
be a good test of cosmological models but will instead provide a tool for 
constraining the physics of galaxy formation.

\end{abstract}
 
 
\section{Introduction}
\bigskip

Over the last three years, the combination of color selection from
deep imaging surveys with spectroscopy from large aperture
telescopes has revealed an abundant population of star-forming
galaxies in the high redshift universe (\cite{steidel96};
\cite{lowenthal97}).  The remarkably rapid growth in surveys of
these ``Lyman-break'' galaxies has created an essentially new
observational field: the study of galaxy clustering at $z>2$.
The first step on this path was the discovery by
Steidel et al. (1998a; hereafter SADGPK) of a sharp spike
in the redshift distribution of Lyman-break galaxies
in a $9^\prime \times 18^\prime$ field, implying that this
galaxy population was already quite strongly clustered at $z=3$.
More recently Giavalisco et al.\ (1998a) have used the angular 
correlation function and the estimated redshift distribution of a
Lyman-break galaxy sample to infer the spatial correlation function,
obtaining a correlation length $r_0=2.1 \pm 0.5\hmpc$ 
(comoving, for $\Omega=1$; $h \equiv H_0/100\;\kms\;{\rm Mpc}^{-1}$).
Adelberger et al.\ (1998, hereafter ASGDPK) have studied the
variance of cell counts in a redshift survey of six
$9^\prime \times 9^\prime$ fields and estimate a somewhat larger
correlation length, $r_0 = 4.0 \pm 1.0\hmpc$ (again for $\Omega=1$).
Other attempts to measure the
galaxy correlation function at high redshift include studies of
angular clustering in the 
Hubble Deep Field (\cite{villumsen97}; \cite{miralles98}) and 
analyses of 3-dimensional clustering of CIV absorption features
in quasar spectra (\cite{quashnock96}; \cite{quashnock97}).

Rapid observational progress begets theoretical interpretations,
and there have been numerous papers offering
predictions of Lyman-break galaxy clustering in different
cosmological models.  Nearly all of these papers in practice
calculate the clustering of dark matter halos, using either
N-body simulations or the approximate analytic methods
developed by Mo \& White (1996) and Kaiser (1984).
The N-body calculations (\cite{bagla98a}b; \cite{jing98b};
\cite{jing98}; \cite{wechsler98}) 
and analytic studies (\cite{mof96}; ASGDPK; \cite{coles98};
\cite{moscardini98}; SADGPK) agree on an essential qualitative
conclusion: if the Lyman-break galaxies in current samples
trace the massive tail of the halo population at $z\sim 3$, then
they are expected to be highly biased tracers of the underlying
mass distribution, as implied by the observed clustering strength.
For this reason, models based on inflation 
and CDM can account reasonably well for the present
clustering data with a variety of choices of cosmological parameters.
ASGDPK turn this argument around to conclude that the
most luminous Lyman-break galaxies must reside in massive halos because
if they were low mass systems in more common halos --- boosted to
high rest-frame UV luminosity by bursts of star formation ---
then they would not exhibit the observed strong clustering.

All of these interpretations rest on the assumption that one can,
at least for purposes of clustering calculations, make a one-to-one
identification between the population of Lyman-break galaxies and
the population of dark matter halos at $z \sim 3$.  The analogous
assumption at $z=0$ is known to be incorrect: typical virialized halos
today represent the common envelopes of galaxy groups or clusters, and
predictions of the galaxy correlation function from CDM N-body
simulations are sensitive to the way one chooses to populate
these halos with galaxies (\cite{gelb94b}).
The one-to-one
assumption may seem more plausible at $z \sim 3$ when the individual
high mass halos have masses that are typical of individual galaxy halos today,
but the galaxies themselves may be smaller at high redshift, and
one might equally well imagine that the relation between massive
halos and galaxy groups is a scaled down version of the one that
exists at $z=0$.  In a hydrodynamic simulation study of the
population of damped Ly$\alpha$ absorbers, Gardner et al.\ (1997,
figure 1) explicitly demonstrate that many dark halos identified
at $z=3$ with density contrast $\rho/{\bar\rho} \sim 200$ contain
more than one galaxy.

Governato et al.\ (1998) have taken a somewhat different approach
to predict the clustering of Lyman-break galaxies.  They use a 
high resolution
N-body simulation to model the formation and clustering of halos,
but they use the semi-analytic galaxy formation prescriptions 
of Cole et al.\ (1994) and Baugh et al.\ (1998) to populate these
halos with galaxies (a similar approach has been taken by Kauffmann
et al. 1998).
This technique substitutes an approximate description of gas cooling,
star formation, and galaxy merging for the one-halo-one-galaxy
assumption.  In principle, this approach could yield significantly
different clustering predictions, although the qualitative conclusion
of Governato et al.\ (1998) --- that the presently detected population
of Lyman-break galaxies should be a highly biased tracer of structure ---
is similar to that of the papers cited earlier (and we will reach
essentially the same conclusion here).

In this paper, we extend our earlier studies of galaxy clustering
in hydrodynamic cosmological simulations (\cite{khw92}; \cite{kwh96},
hereafter KWH) to an investigation of high redshift galaxy clustering
in five variants of the CDM scenario (\cite{peebles82}; 
\cite{blumenthal84}).  Two of the simulations
(the SCDM and CCDM models described below) are the ones analyzed
in KWH and Weinberg, Hernquist, \& Katz (1997, hereafter WHK), 
but we did not present galaxy
correlation functions in those papers because even that short time
ago it seemed that the clustering of $z=3$ galaxies would not be
accessible to observation for many years.  The advantage of full
hydrodynamic simulations over the semi-analytic galaxy formation
method is that the crucial processes of dissipation and merging
are treated much more realistically, without the idealizing
approximations of spherical symmetry, isothermal profiles, 
and parameterized merging rates.  Galaxy identification is
straightforward and robust in high resolution hydrodynamic
simulations because the dissipated objects stand out
distinctly from the background and from each other, with overdensities
of $10^3-10^6$ or even higher.  The physics of galactic scale
star formation and feedback is an uncertainty in both the semi-analytic
and the hydrodynamic simulation approaches, but at least within
the range of assumptions explored by KWH the inferred galaxy population
is insensitive to details of the star formation prescription.
Indeed, one finds nearly identical galaxy populations in simulations
that include star formation and feedback and in simulations that
have no star formation and identify galaxies directly as clumps
of cold, dense gas (WHK, figure 10; \cite{pearce98}).
The galaxy population is also completely insensitive to the presence
of a photoionizing background in the mass range that we are
presently able to resolve (WHK).

Since the observed-frame optical (rest-frame UV) luminosity of a $z=3$
galaxy depends mainly on its instantaneous star formation rate,
the identification of which of our simulated galaxies would make
it into an observed Lyman-break galaxy sample {\it is} sensitive
to our handling of star formation.  Through most of this paper,
we will make the simplifying assumption that a simulated galaxy's
UV luminosity is a monotonically increasing function of its
baryonic mass, but in \S 4 we will briefly discuss the
possibility that episodic star formation introduces large scatter
into this relationship.  We will address issues related to
the luminosity function of Lyman-break galaxies and the cosmic
star formation history in a separate paper (for preliminary
results see Weinberg, Katz, \& Hernquist 1998, hereafter WKH).

The disadvantage of the numerical hydrodynamic approach to this problem
is that computational expense forces us to adopt relatively small
simulation volumes;
in the case of the models described here,
periodic cubes 11.111$h^{-1}$ comoving Mpc on a side.
This small volume affects our results in several ways.
First, our simulation volume would contain only a handful of
objects at the space density of the SADGPK sample, so we will mainly
examine the clustering properties of the more numerous galaxies
fainter than the magnitude limits of existing {\it spectroscopic}
samples of high-$z$ objects.  Second, we have only a single realization
of each model, and there might be significant
statistical fluctuations in the clustering from one such volume
to another.  Third, the absence of waves larger than our box
causes us to systematically underestimate the true amplitude of
clustering.  While the nonlinear scale of the dark matter is generally
much smaller than the box at $z \geq 2$, the galaxies exhibit
stronger clustering than the dark matter.  Extreme
regions like a proto-Coma cluster or a proto-Bo\"otes void would be
as large as our simulation volume (and the structure discovered
by SADGPK might be roughly such a proto-Coma region).  Our approach
to estimating these finite volume effects (in \S 3.4 below)
will be to assume
that our hydrodynamic simulations correctly compute the {\it relative}
amplitudes of the galaxy and dark matter correlation functions, allowing us to
calibrate these bias factors in the hydrodynamic
simulations and apply them to larger volume, lower resolution N-body
simulations in order to estimate corrected correlation lengths.
Our treatment still leaves us with statistical and systematic 
uncertainties in the clustering predictions, which are probably
comparable in magnitude to the statistical and systematic uncertainties
in the current measurements.  We can hope for significant progress
on both the theoretical and observational fronts over the next few years.

\section{Methods and Models}
\bigskip

We perform our simulations using TreeSPH (\cite{hk89}), a code that
unites smoothed particle hydrodynamics (SPH; \cite{lucy77};
\cite{gingold77}) with the hierarchical tree method for computing 
gravitational forces (\cite{bh86}; \cite{h87}).  Dark matter, stars,
and gas are all represented by particles; collisionless material is
influenced only by gravity, while gas is subject to gravitational
forces, pressure gradients, and shocks.  We include the effects of
radiative cooling, assuming primordial abundances, and Compton
cooling.  Ionization and heat input from a UV radiation background are
incorporated, assuming optically thin gas, using the radiation field
computed by Haardt \& Madau (1996).  
We use a simple prescription to turn cold, dense gas into collisionless
``star'' particles.  The technique and its computational implementation
are described in detail by KWH.  In brief, gas becomes ``eligible'' to
form stars if it has a physical density corresponding to
$n_{\rm H} > 0.1\; \cm^{-3}$ and an overdensity $\rho/{\bar\rho}>56.7$
(equivalent to that at the virial radius of an isothermal sphere).
The gas must also reside in a convergent flow and be locally Jeans 
unstable, although the density criteria themselves are usually sufficient
to ensure this.  
Eligible gas is converted to stars at a rate 
$d{\rm ln}\rho_g/dt = -c_\star/t_g$, where $t_g$ is the maximum of
the dynamical time and the cooling time.  We use $c_\star=0.1$ for
the simulations here, but KWH show that the simulated galaxy
population is insensitive to an order-of-magnitude change in
$c_\star$, basically because the star formation rate is forced
into approximate balance with the rate at which gas cools and
condenses out of the hot halo.  
When star formation occurs,
supernova heating is added to the surrounding gas assuming a standard
IMF from $0.1$ to $100 {\rm M}_\odot$ and that stars above $8 {\rm
M}_\odot$ become supernovae.  Each supernova adds $10^{51}$ ergs of
thermal energy to the system.  

Because it uses a Lagrangian hydrodynamics algorithm and individual
particle time steps, TreeSPH makes possible simulations with the
enormous dynamic range needed to study galaxy formation in a cosmological
context.  In SPH, gas properties are computed by averaging
or ``smoothing'' over a fixed number of neighboring particles,
32 in the calculations here.
When matter is distributed homogeneously, all
particles have similar smoothing volumes.  However, smoothing lengths
in TreeSPH are allowed to decrease in collapsing regions, in
proportion to the interparticle separation, thus increasing the
spatial resolution in precisely those regions where a high dynamic
range is needed.  In underdense regions, the smoothing lengths are
larger, but this is physically reasonable because the gas distribution
{\it is} smoother in these regions, requiring fewer particles for an
accurate representation.  TreeSPH allows particles to have individual
time steps according to their physical state, so that the pace of the
overall computation is not driven by the small fraction of particles
requiring the smallest time steps.  In the calculations described
here, the largest allowed timestep is $3.25 \times 10^6$ years.  The
smallest allowed timestep is 16 times smaller, or $2.03 \times 10^5$
years.  If the Courant condition would demand a timestep smaller
than this minimum, we degrade the gas resolution (increase the SPH
smoothing length) so that the Courant condition is satisfied.
The timestep criteria are detailed further in KWH and Quinn et al.\ (1998);
we set the tolerance parameter $\epsilon_{\rm tol} = 0.4$. 

Here we present the results of five simulations,
one each of five different
cosmological models.  All these simulations, of periodic cubes that are
$11.111\hmpc$ on a side, use
$2 \times 64^3$ particles and are evolved to $z=2$.  Each has a nominal
gas mass resolution (32 gas particles) of $4.7 \times 10^{9}
(\Omega_b/0.05)\msun$ and a spatial resolution (gravitational
softening length, in physical units) of $3(1 + z)^{-1} \hkpc$
(equivalent Plummer softening).  
We adopt Walker et al.'s (1991) estimate of the baryon density,
$\Omega_b=0.0125h^{-2}$.  We would not expect a higher value
of $\Omega_b$ ({\it e.g.} \cite{burles97}, 1998; Rauch et al. 1997) 
to alter our clustering
results, but it would increase the baryonic mass (and presumably
the luminosity) of galaxies at fixed space density.
We adopt the same Fourier phases for the initial conditions
of each model to minimize the impact of statistical fluctuations
in our finite simulation volume on the comparison between models.

Our first model is ``standard'' CDM (SCDM), with $\Omega=1$, $h=0.5$.
The power spectrum is normalized so that the rms amplitude of mass
fluctuations in $8\hmpc$ spheres, linearly extrapolated to $z=0$, is
$\sigma_{8}=0.7$.  This normalization is consistent with 
(but somewhat higher than) that
advocated by White, Efstathiou \& Frenk (1993) to match the observed
masses of rich galaxy clusters.  However, it is inconsistent with the
normalization implied by the COBE-DMR experiment.  Our second model
(CCDM) is identical to the first one except that $\sigma_{8}=1.2$,
consistent with the 4-year COBE data (\cite{bennett96}) but inconsistent
with the $z=0$ cluster population.  The third model, OCDM,
assumes an open universe with $\Omega=0.4$, $h=0.65$, and
$\sigma_8=0.75$.  It is 
COBE-normalized (\cite{ratra97}) and produces an acceptable
cluster mass function (\cite{cwfr97}).
The fourth model, a nonzero-$\Lambda$ CDM model (LCDM),
has $\Omega=0.4$, $\Lambda=0.6$, $h=0.65$, and an inflationary 
fluctuation spectrum tilted ($n=0.93$) so that it simultaneously
matches the COBE and cluster constraints.
The fifth model (TCDM) is an $\Omega=1$, $h=0.5$ model that also has a
tilted inflationary spectrum ($n=0.8$) chosen so that it matches
COBE and cluster constraints.
We list the parameters of all the models in Table 1.  We also include
a lower resolution simulation of the SCDM model evolved to $z = 0$
(from KWH).  This simulation has 1/2 the spatial resolution
and 1/8th the mass resolution of the other simulations, and it
does not incorporate a photoionizing background.

\begin{table}
\caption{Models} \label{tbl-1}
\begin{center}\scriptsize
\begin{tabular}{crrrrrrrrrrr}
Model & $\Omega$ & $\Lambda$ & $H_0$ & $\Omega_b$ & $\sigma_{8}$ & $n$ \\
\tableline
SCDM  &     1.0  &    0.0  &     50  &    0.05  &    0.7   &   1.0 \\
CCDM  &     1.0  &    0.0  &     50  &    0.05  &    1.2   &   1.0 \\
OCDM  &     0.4  &    0.0  &     65  &    0.03  &    0.75  &   1.0 \\
LCDM  &     0.4  &    0.6  &     65  &    0.03  &    0.8   &   0.93 \\
TCDM  &     1.0  &    0.0  &     50  &    0.05  &    0.54  &   0.80 \\

\end{tabular}
\end{center}
\end{table}

As we discuss below, the interplay between the dynamics of the gas and
the microphysical processes that influence its thermal state
leads to the development of a three phase medium.  In what follows we
are particularly interested in the cold, dense component, which we
identify with forming galaxies.  These clumps of gas are at
sufficiently high overdensity that stars form readily in them.
Because the knots of cold gas particles and star particles are compact
and distinct, it is easy to group them into simulated ``galaxies.''
The specific algorithm that we adopt is SKID
(Spline Kernel Interpolative DENMAX; KWH, 
http://www-hpcc.astro.washington.edu/tools/SKID/),
which identifies galaxies
as gravitationally bound groups of cold gas and star particles that
are associated with a common density maximum.
The approach is inspired by Gelb \&
Bertschinger's (1994a) DENMAX algorithm, but instead of defining
gradients of the density field on an Eulerian mesh, it uses the SPH
smoothing kernel to calculate Lagrangian density gradients.  Only star
particles and gas particles that have $\rho/{\bar \rho} \geq 1000$
and $T \leq 30,000\;\K$ are eligible to be galaxy particles.
We require a SKID group to contain at least eight particles in order 
to count it as a galaxy.  While eight may seem a dangerously low
number, the 32-particle averaging in SPH ensures that any collection
of eight cold, high density particles must reside in a significantly
more massive overdense background.

It is useful to express the resolution limit of the simulations
in terms of the minimum
circular velocity of halos in which we can resolve galaxies.
Gardner et al.\ (1997) examined all halos in our SCDM simulation 
and found that only those with $v_c \geq 100\;\kms$ (measured at
the virial radius) contained dense, cold objects corresponding
to galaxies.  We have applied the same analysis to our other models
and find that the limit is $v_c \geq 100\;\kms$ in the $\Omega=1$
models and $v_c\geq 70\;\kms$ in the low-$\Omega$ models (LCDM, OCDM).
Although we cannot be certain that our results for galaxies above
this mass limit do not suffer from finite resolution effects, the
tests of Owen, Weinberg, \& Villumsen (1998) for self-similar evolution
in hydrodynamic simulations with scale-free initial conditions
suggest that once a simulation has enough resolution to follow the
formation of radiatively cooled gas clumps it obtains robust results
for the subsequent growth and mergers of these objects.

\section{Results}
\bigskip

\subsection{The High Redshift Galaxy Populations}
\bigskip

As discussed in KWH and WKH, the evolution of the dark matter in the SPH
simulations is similar to that in collisionless N-body simulations.
The gas component, with a mass fraction that ranges
from 0.050 to 0.074 in the different models, has little dynamical
effect on the dark matter except in small regions surrounding the
densest concentrations of cold gas.  Gravitational instability 
produces a network of filaments and clumps containing both dark matter
and gas, with the most prominent clumps residing at the intersections
of filaments.  When gas falls into a dark matter potential well,
it is shock heated to roughly the virial temperature,
converting its kinetic infall energy into thermal
energy.  The gas is supported against further collapse by thermal pressure.
However, the densest gas is able to dissipate
energy through atomic and Compton cooling processes, so it loses
pressure support and settles into highly overdense knots within the
dark halos.  These clumps are so dense that they remain cold even
when mergers occur or when they fall into larger halos.  They are also
much more resistant to tidal disruption than
the relatively puffy dark matter halos that form in a purely
dissipationless simulation, so the clumps can survive as distinct
entities even when their parent halos merge.

At high redshift the gas in SPH simulations resides in 
three main components: low
density, highly photoionized gas with $\rho/\bar{\rho} \lta 10$ and $T \lta
10^5$ K, shock heated gas with typical overdensity $\rho/\bar{\rho}
\sim 10$--$10^4$ and $T \sim 10^5$--$10^7$ K, and radiatively cooled,
dense gas with $\rho/\bar{\rho} \gta 1000$ and $T \sim 10^4$ K
(see KWH, figure 3).  
The first component gives rise to the \lya forest, the second component is
associated with hot gas halos of galaxies, galaxy groups, and clusters,
and the third component is associated with damped
\lya systems and high redshift galaxies.
In simulations with star formation, gas from this third component
is converted steadily into stars.

Figure 1 shows the spatial distribution of galaxies at $z = 3$ for the
five models, with the galaxies identified by the procedure
described in \S 2.  Each galaxy is represented by a circle whose
area is proportional to its baryonic mass (cold gas plus stars).  
The concentration of galaxies in the middle right
of the panels becomes a small cluster of galaxies by $z = 0$, at
least in the SCDM model (KWH; WKH).
For SCDM, CCDM, and TCDM the angular size of each comoving panel at $z=3$
is 12.7 arc minutes, and the redshift depth is $\Delta z = 0.0297$.  For
OCDM, the angular size is 9.5 arc minutes with $\Delta z =
0.0220$, and for LCDM model the angular size is 9.3 arc minutes
with $\Delta z = 0.0190$.  The redshift depths of the simulation volumes
are therefore small compared to the survey volume of SADGPK, who
observed a field $17.64^\prime \times 8.74^\prime$ and found 
Lyman-break galaxies between $z = 2.7$ and $z = 3.4$, i.e., $\Delta z = 0.7$.  

\begin{figure*}
\centerline{
\epsfxsize=4.5truein
\epsfbox[65 0 550 740]{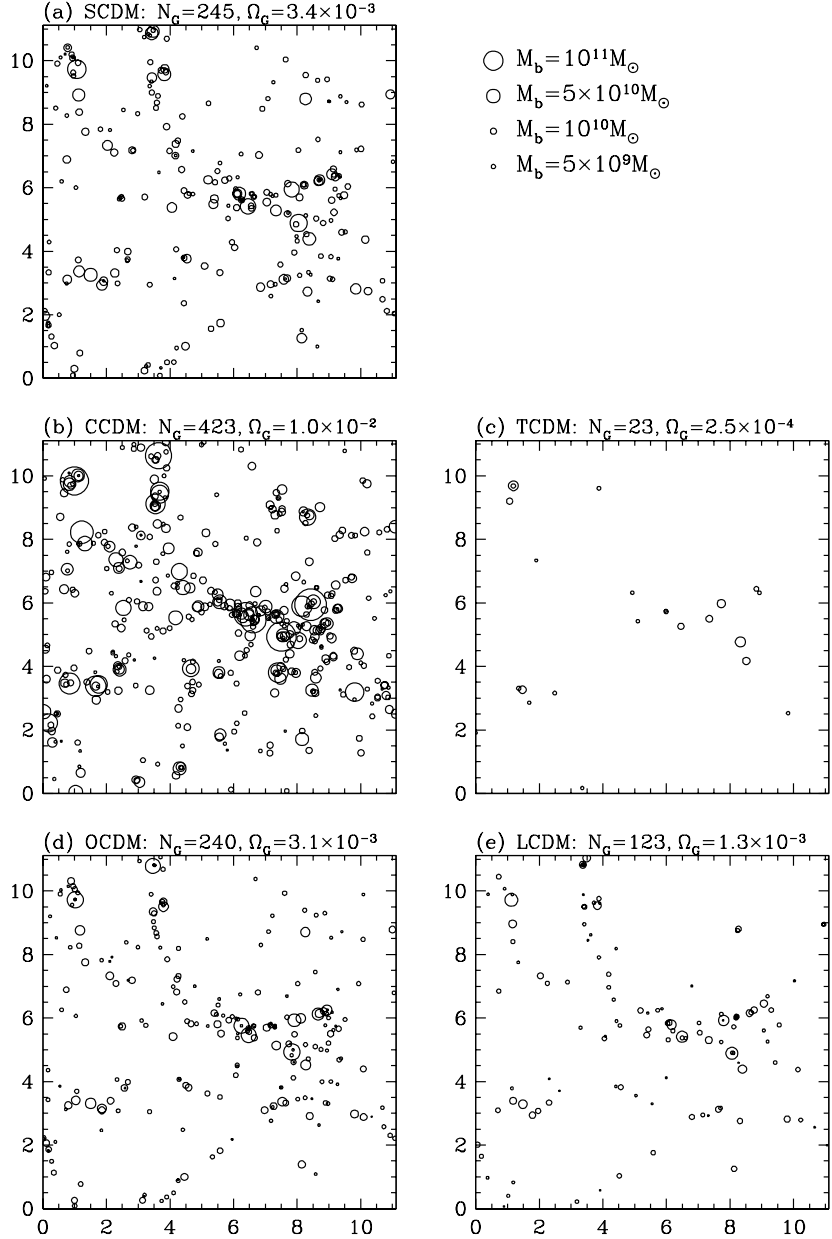}
}
\caption{              
\label{fig:circle}
The distribution of ``galaxies'' in each of the models at $z =
3$. Each galaxy is represented by a circle whose area is proportional
to its baryonic mass (cold gas plus stars).  The total number of
galaxies and $\Omega_g$, the fraction of the critical density in
galactic baryonic mass, is also indicated for each model.  The axes
are marked in comoving $h^{-1}$ Mpc.
}
\end{figure*}

\begin{figure*}
\centerline{
\epsfxsize=4.5truein
\epsfbox[65 0 550 740]{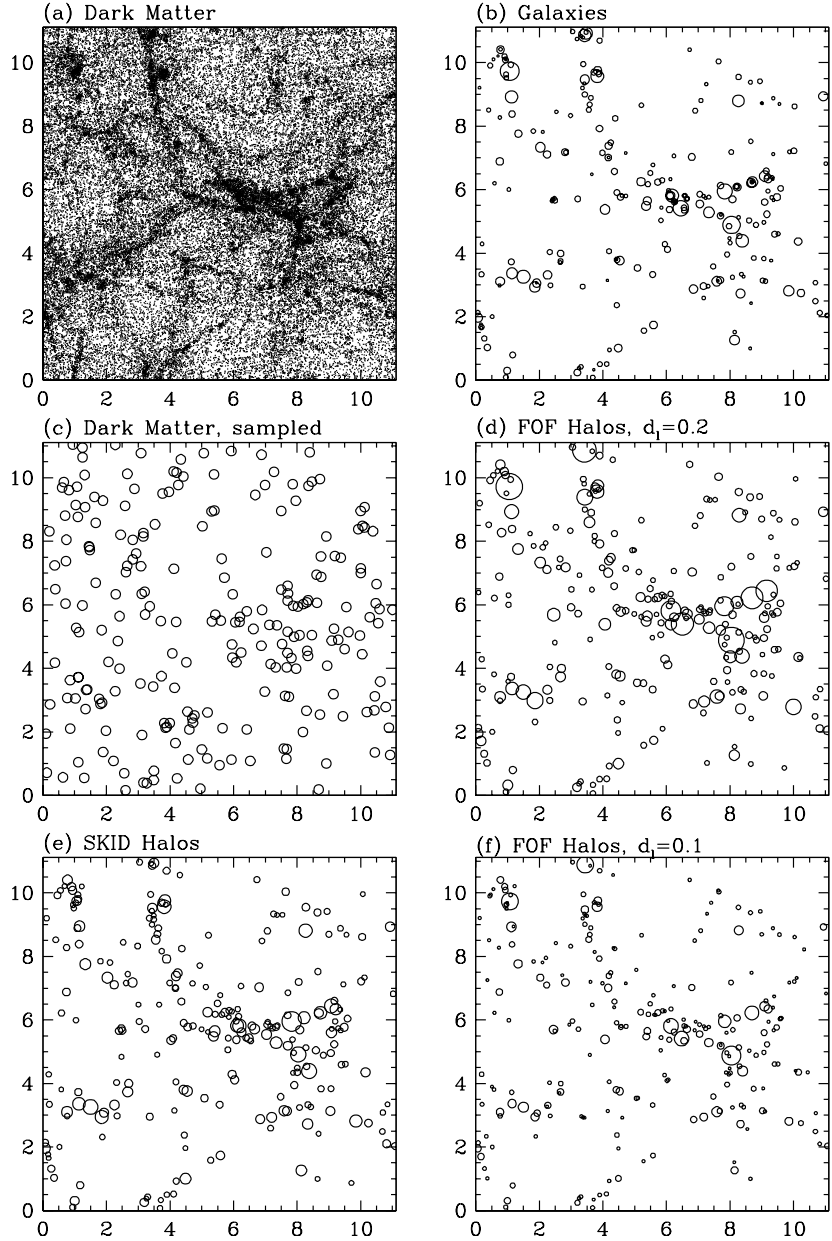}
}
\caption{              
\label{fig:halo}
Distributions of dark matter, galaxies, and halos in the SCDM model at $z=3$.  
({\it a})
Dark matter particles (randomly selected subset of 1 in 4) from a
purely gravitational, N-body simulation.
({\it b})
SPH galaxies, repeated from Fig.~\ref{fig:circle}a.
({\it c})
Random subset of 245 dark matter particles from (a), equal to the number
of SPH galaxies in (b).
({\it d})
The 245 most massive halos from the N-body simulation, identified
by the friends-of-friends (FOF) algorithm with linking parameter
$d_l=0.2$.
({\it e})
Same as (d) for SKID halos.
({\it f})
Same as (d) for linking parameter $d_l=0.1$.
}
\end{figure*}

\begin{figure*}
\centerline{
\epsfxsize=4.5truein
\epsfbox[65 0 550 740]{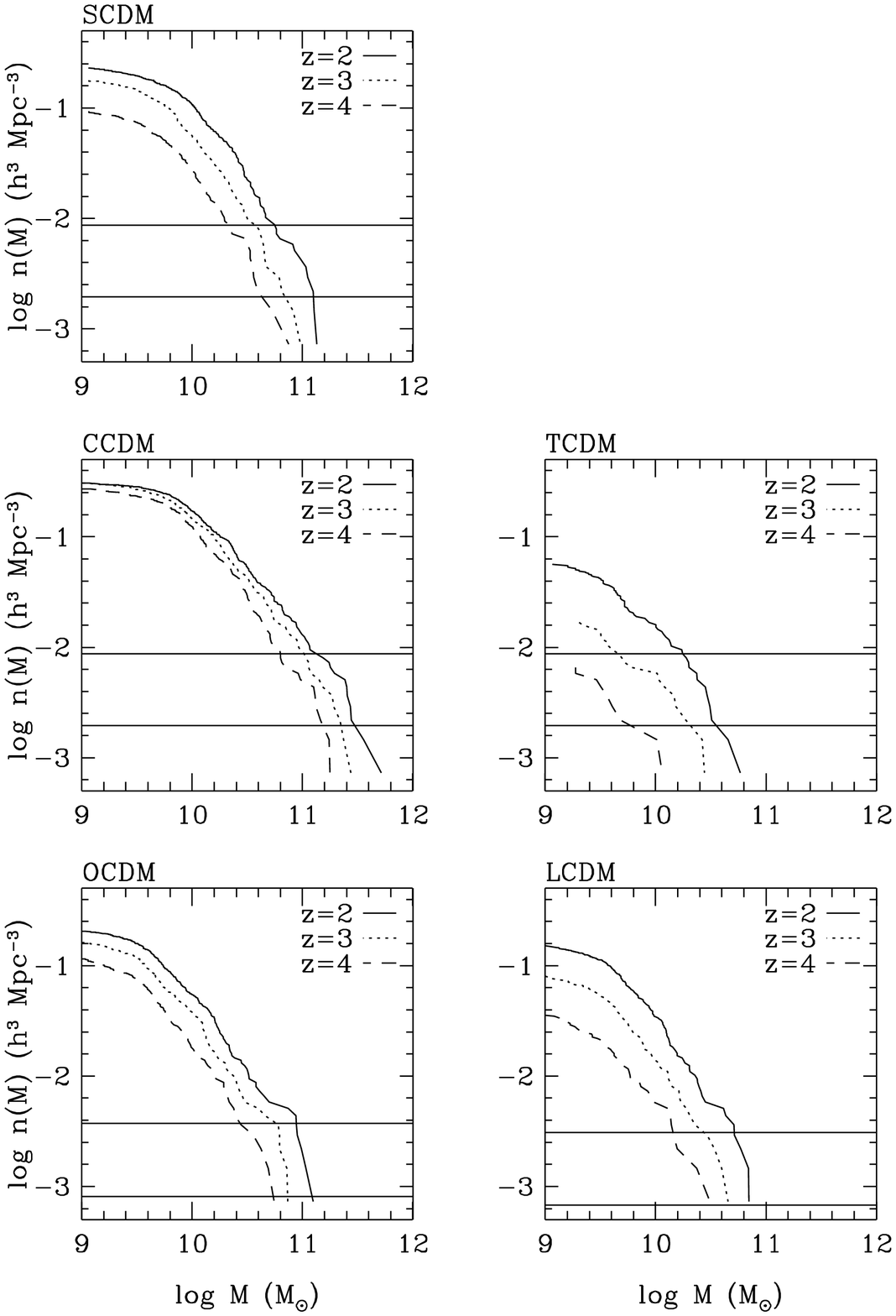}
}
\caption{              
\label{fig:massfun}
Cumulative baryonic mass function in the different models at the
indicated redshifts: $n(M)$ is the comoving number density of galaxies
whose baryonic mass exceeds $M$.  The solid horizontal lines mark
the space density of Lyman-break galaxies in the SADGPK sample
(lower line) and the Lowenthal et al.\ (1997) sample (upper line).
}
\end{figure*}

The more massive galaxies are predominantly stellar and the less
massive ones are more gas-rich, though this trend could be partly an
artifact of limited resolution, which influences small galaxies more
than larger ones.  By $z=3$, the largest galaxies already have
baryonic masses of $2.8 \times 10^{11} \msun$ in the CCDM model, but they
have reached a baryonic mass of only $2.8 \times 10^{10} \msun$ in
the TCDM model.  This difference directly reflects the higher
amplitude of mass fluctuations in the CCDM model; the remaining
models have intermediate fluctuation amplitudes and intermediate
galaxy masses.

The upper left panel of Figure~\ref{fig:halo} shows the dark 
matter particle distribution from a pure N-body simulation of the
SCDM model, started from the same initial conditions as the SPH 
simulation and evolved with the same simulation code and numerical parameters.
The dark matter distribution of the SPH simulation itself
would be essentially indistinguishable at the level of detail visible
on this plot.  Figure~\ref{fig:halo}b shows the SPH galaxy distribution,
repeated from Figure~\ref{fig:circle}.  Comparing these two panels shows
that the galaxy distribution traces the prominent features of the
underlying mass distribution.  However, randomly sampling the dark 
matter particles to the space density of the galaxies produces a 
distribution that is clearly less structured (Fig.~\ref{fig:halo}c).
This comparison demonstrates visually the point that we will quantify
in the next section: the high-redshift galaxies in CDM models
are highly biased tracers of the underlying structure, concentrated
in clumps and filaments and avoiding underdense regions.

The remaining panels of Figure~\ref{fig:halo} show the populations
of dark matter halos in the N-body simulation, identified by different
algorithms.  For Figure~\ref{fig:halo}d we adopt the most widely
used procedure for defining halos, the ``friends-of-friends'' (FOF)
algorithm with linking parameter $d_l=0.2$ in units of the
mean interparticle separation (see, e.g., \cite{frenk88}; the 
implementation used here is available at
http://www-hpcc.astro.washington.edu/tools/FOF/).
We select the 245 most massive FOF halos, so that the number of
halos is equal to the number of SPH galaxies in the panel above.
The mass scale for determining the sizes of symbols has been
shifted by a factor $\Omega/\Omega_b=20$.  If each FOF
halo contained a single galaxy with a baryon-to-dark-matter
ratio equal to the universal value, Figures~\ref{fig:halo}b 
and~\ref{fig:halo}d  would be
be identical.  The bottom panels of Figure~\ref{fig:halo} show the
halo populations identified by FOF with a shorter linking
length, $d_l=0.1$, corresponding to a higher effective threshold
density, and by the SKID algorithm (with a smoothing scale of 64
particles and an overdensity threshold of 50).

Figure~\ref{fig:halo} shows that the dark matter halos are also
biased tracers of structure at $z=3$ --- their space distribution
resembles the galaxy distribution much more than it resembles the
dark matter distribution.  However, while the halo and galaxy distributions
give a similar overall picture, there is clearly not a one-to-one
match between halos and galaxies in the high density regions
that dominate statistical clustering measures, and the halo population
itself depends on the specific identification algorithm.
It is these differences between halos and galaxies
and among halo identification algorithms themselves that motivates
the use of hydrodynamic simulations to study high-redshift galaxy
clustering.  As already noted, the galaxy population in high resolution
SPH simulations is insensitive to ``microphysical'' assumptions
about star formation, feedback, and photoionization, at least within
the range tested by KWH, WHK, and Pearce (1998).
Furthermore, because dissipation produces highly overdense objects,
there is essentially no ambiguity in grouping cold gas and star
particles into galaxies: N-body halos overlap, but SPH galaxies do not.
While halo-based models can provide a useful qualitative guide to 
expectations of high-redshift structure, collisionless N-body simulations
do not include the dissipational physics needed to obtain robust
and accurate quantitative predictions of galaxy clustering at any
redshift.  We must also acknowledge, however, that the small
volume of our current SPH simulations is also a significant
limitation, and that larger and more ambitious simulations will
be needed to take advantage of improvements in observational constraints.

Figure~\ref{fig:massfun} plots the cumulative baryonic mass function of the
galaxies in the five models at $z = 2$, 3, and 4. The high end of 
the mass function shifts to the right 
in each of the models as more gas condenses into galaxies and
small galaxies merge.  In the highest amplitude model, CCDM,
the low mass end ceases to evolve as smaller galaxies merge into
larger ones and are not replaced. 
Limited mass resolution almost certainly causes the
turnover in all models at the low mass end, and it 
is likely responsible for the lack of
late evolution in the CCDM model at these masses.

As one might expect, given the relative power on galactic scales, the
CCDM model has the most galaxies at $z = 2$, 3, and 4, followed by OCDM,
SCDM, LCDM, TCDM. 
The cumulative mass function evolves fastest in the models
with the weakest high-redshift mass fluctuations, LCDM and TCDM.
The horizontal lines mark the space densities of systems detected
in the Lyman-break surveys of SADGPK and Lowenthal et al.\ (1997).
SADGPK survey a volume of $\sim 30,400
\hmpccc$ ($\sim 72,900 \hmpccc$; $88,000 \hmpccc$) for $\Omega = 1$
($\Omega = 0.4$; $\Omega = 0.4$, $\Lambda = 0.6$) and 
obtain spectroscopic confirmation of 
59 Lyman-break galaxies, implying a space density of
$1.94 \times 10^{-3} \invhmpccc$ ($8.10 \times
10^{-4} \invhmpccc$; $6.70 \times 10^{-4} \invhmpccc$). Lowenthal et
al. (1997) observe to a fainter magnitude limit and find a number
density of $8.8 \times 10^{-3} \invhmpccc$ ($3.7 \times 10^{-3}
\invhmpccc$; $3.1 \times 10^{-3} \invhmpccc$).  On average, we would
expect 2.7 (1.1; 0.9) galaxies at the SADGPK density and 12.1 (5.1;
4.3) galaxies at the Lowenthal et al. density in our simulation
volume.  The more extensive ASGDPK sample has a space density
similar to that of SADGPK.
If our CDM models are representative of the real universe,
then the objects in existing spectroscopic samples of
Lyman-break galaxies represent just the luminous tip of the
full population of galaxies at these redshifts, presumably corresponding
to the largest few circles in the panels of Figure~\ref{fig:circle}.

\subsection{Galaxy Correlation Functions}

\begin{figure}
\centerline{
\epsfxsize=4.5truein
\epsfbox[65 0 550 740]{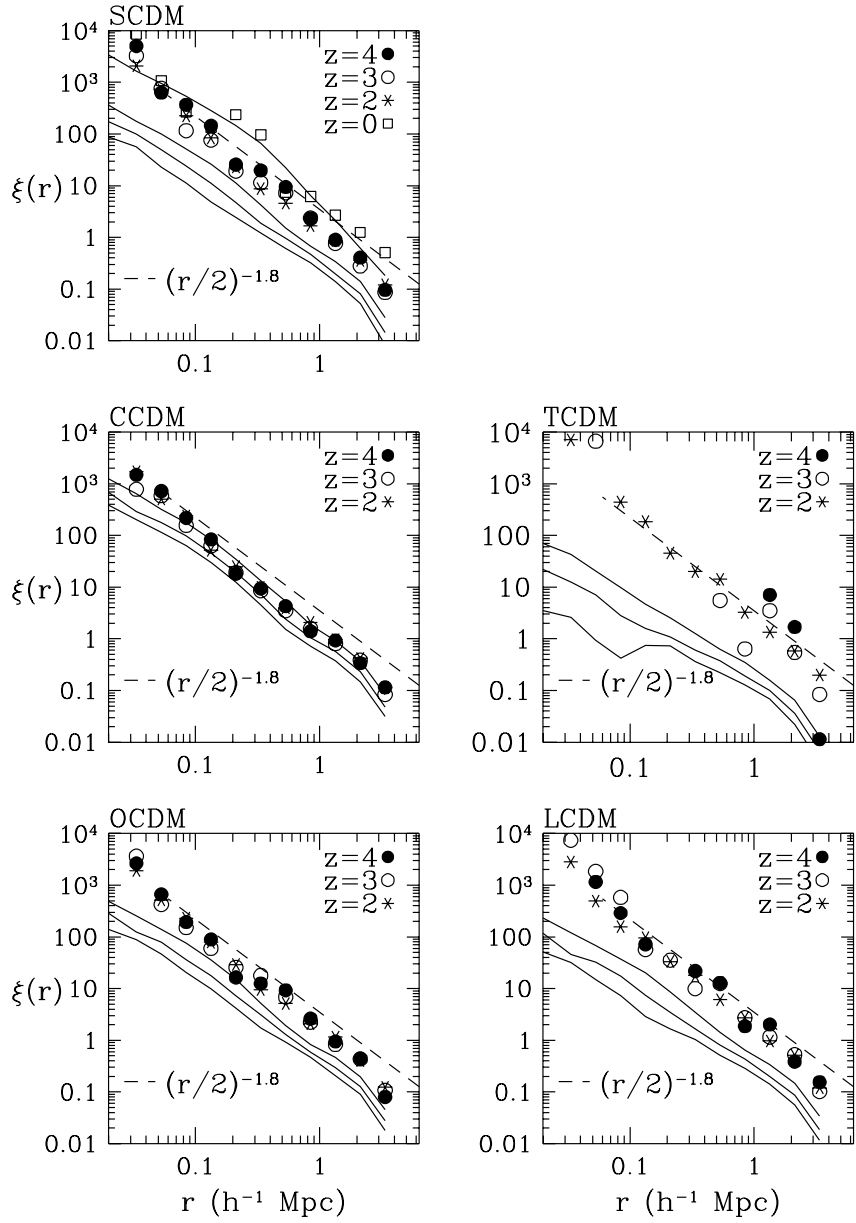}
}
\caption{              
\label{fig:xi1}
Galaxy correlation functions, in comoving distance units.
Symbols show the correlation functions of SPH galaxies at the
indicated redshifts.
Solid lines show the dark matter correlation functions at the same
redshifts (increasing in amplitude with decreasing redshift).
The dashed line is a power law with a slope of $-1.8$
and a correlation length of $2.0 \hmpc$.
}
\end{figure}

Figure~\ref{fig:xi1} presents the principal results of this paper,
the galaxy correlation functions in our five cosmological models at
$z = 2$, 3, and 4. We calculate the correlation function
for all the galaxies in the simulation volume.  
As discussed in \S 2 above, the effective circular velocity limits
of the simulations are $v_c \geq 100\kms$ in the SCDM, CCDM, and TCDM
models and $v_c \geq 70\kms$ in the OCDM and LCDM models.
We also calculate the correlation function at $z = 0$
for the SCDM model using the lower resolution simulation mentioned
in \S 2, which has an effective circular velocity limit $v_c \geq 200 \kms$. 
The TCDM results are noisy except
at $z=2$ because this model does not form many galaxies above our
resolution limit at higher redshifts.
Points show the galaxy correlation function at various redshifts
and solid lines show the dark matter correlation functions, which
increase monotonically in time as mass fluctuations grow.
For comparison purposes, the dashed line in each panel shows a 
power law correlation function,
\begin{equation}
\xi(r) = (r/r_0)^{\gamma},
\label{eqn:xi}
\end{equation}
with a slope $\gamma = -1.8$ and a correlation length $r_0 = 2.0
\hmpc$ (comoving units are used throughout). 
In all cases with adequate statistics, the galaxy correlation function
is well described by a power law over 2 decades in $r$ (roughly
four decades in $\xi$), though the slope is usually closer
to $-2.0$ than to $-1.8$.

Figure~\ref{fig:xi1} exhibits several striking features.
In all models, the galaxy correlation function shows little
or no evolution between $z=4$ and $z=2$, although the dark matter
correlation function grows steadily stronger over this period.
The galaxy correlations are more pronounced
than the dark matter correlations
in virtually all cases, sometimes by a factor of 10 or more.
This strong bias exists despite our fairly low circular velocity
threshold and correspondingly high galaxy space density.
Perhaps most remarkably, there is very little difference in the
galaxy correlation function from one model to another, even though
these models have different cosmological parameters and different
mass fluctuation amplitudes.  To the extent that there are 
differences between models, it is the model with the weakest mass 
fluctuations (TCDM) that has the strongest galaxy correlations at $z=2$,
and it is the model with the strongest mass correlations (CCDM)
that has the weakest galaxy correlations.  Within an individual
model, there is often a slight drop in the galaxy correlation 
amplitude from $z=4$ to $z=2$.

The correlation function results quantify the visual
impressions of Figures~\ref{fig:circle} and~\ref{fig:halo}: 
the high redshift galaxies form in ``special'' locations that 
trace the skeleton of structure
in the mass distribution.  This skeleton is essentially the same
from one model to another, even though its contrast in the mass
distribution depends on the amplitude of mass fluctuations in the model.
Between $z=4$ and $z=2$, an increasing amount of dark matter flows into
this skeleton, but this does not substantially increase the 
amplitude of galaxy clustering because the galaxies are already
there.  Only at low redshifts does the gravitational clustering
of the dark matter begin to carry the galaxies into higher contrast
structures and increase their correlation function, as seen in
the $z=2$ to $z=0$ evolution of the SCDM model.

\begin{figure}
\centerline{
\epsfxsize=4.5truein
\epsfbox[50 470 555 725]{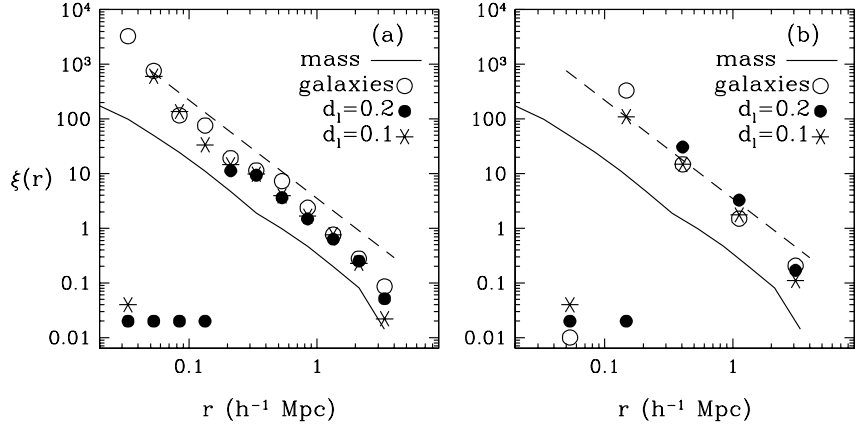}
}
\caption{              
\label{fig:xihalo}
Correlation functions of dark matter (solid line), SPH galaxies (open 
circles), and FOF halos identified with linking parameter $d_l=0.2$
(filled circles) and $d_l=0.1$ (asterisks), for the SCDM model at $z=3$.
Bins at small separations containing no pairs are indicated by points near
the horizontal axis.  The dashed line shows a power law
$\xi(r)=(r/2)^{-1.8}$.
({\it a}) Comparison of all (245) SPH galaxies to the most massive 245
halos.
({\it b}) Comparison of the 30 most massive SPH galaxies to the 30 most
massive halos.
}
\end{figure}

Figure~\ref{fig:xihalo} compares the galaxy and mass correlation functions 
measured from the SCDM SPH simulation at $z=3$
to correlation functions of FOF halos
from the SCDM N-body simulation at the same redshift.
For Figure~\ref{fig:halo}a, we select the 245 most massive halos,
so that the space density of the halos is equal to the space
density of the SPH galaxies.
The FOF halo identification method effectively precludes halos from
being very close neighbors, so there are no pairs (implying $\xi = -1$)
at $r<0.2\hmpc$ for linking parameter $d_l=0.2$.
At larger separations the halo correlation function is slightly
weaker than the galaxy correlation function, but it is substantially
stronger than the dark matter correlation function, as one
would expect based on Figure~\ref{fig:halo}.
Halos selected with $d_l=0.1$ have a similar correlation function,
except that $\xi(r)$ continues to rise in to smaller separations.
In both cases, the halos are positively biased, despite the relatively
high comoving space density of 0.18 $h^3 {\rm Mpc}^{-3}$.
Figure~\ref{fig:xihalo}b compares the correlation function of the
30 most massive SPH galaxies to the 30 most massive FOF halos,
with $\xi(r)$ averaged over larger bins to reduce noise.
The clustering of the galaxies increases slightly with the higher
mass threshold (a point we will return to later), but the clustering
of the $d_l=0.2$ halos increases more, so that at intermediate
separations they are now more strongly clustered than the galaxies.

For nearly all cases in Figure~\ref{fig:xi1}, the galaxy correlation
function is close to a power law in the range
$0.04\hmpc < r < 4\hmpc$.  Tables 2 and 3 list the parameters
$r_0$ and $\gamma$, respectively, of power law fits to these correlation
functions, where $r_0$ is the correlation length and $\gamma$
the power law index (see equation~[\ref{eqn:xi}]).
We perform the fits in log space over the above radial range,
using Poisson counting errors to estimate the error in $\xi(r)$
in each radial bin.  
At $z=3$ the correlation lengths range from $r_0=1.13\hmpc$
to $r_0=1.64\hmpc$, and the power law slopes range from $\gamma=-1.97$
to $\gamma=-2.46$.  Note, however, that our finite box size
causes us to systematically underestimate $r_0$ and to 
systematically overestimate $|\gamma|$, a point that we will return
to in \S 3.4 below.  The uncertainties listed in Tables~2 and~3
are the formal uncertainties of the parameter fits, but they
do not include the statistical uncertainty that arises because we have
only a single realization of each model.  We cannot assess the
magnitude of this uncertainty until we perform more simulations.
However, these are reasonable error bars to use when comparing
one model to another, since we employ the same initial Fourier phases
in each simulation.

\begin{table}
\caption{Correlation Lengths, $r_0$, in $\hmpc$} \label{tbl-2}
\begin{center}\scriptsize
\begin{tabular}{crrrrrrrrrrr}
Model & $z=4$ & $z=3$ & $z=2$ & $z=1$ & $z=0$ \\
\tableline
SCDM  & $1.29 \pm 0.03$  & $1.16 \pm 0.02$  & $1.20 \pm 0.01$ & $1.72 \pm 0.10$ & $2.35 \pm 0.08$ \\
CCDM  & $1.17 \pm 0.01$  & $1.13 \pm 0.01$ & $1.23 \pm 0.01$ &     -    &     -   \\
OCDM  & $1.28 \pm 0.02$  & $1.27 \pm 0.02$ & $1.28 \pm 0.01$ &     -    &     -   \\
LCDM  & $1.51 \pm 0.07$  & $1.41 \pm 0.03$ & $1.35 \pm 0.02$ &     -    &     -   \\
TCDM  & $2.04 \pm 0.34$  & $1.64 \pm 0.15$ & $1.59 \pm 0.06$ &     -    &     -   \\

\end{tabular}
\end{center}
\end{table}

\begin{table}
\caption{Power law slope, $\gamma$, of the correlation functions} \label{tbl-3}
\begin{center}\scriptsize
\begin{tabular}{crrrrrrrrrrr}
Model & $z=4$ & $z=3$ & $z=2$ & $z=1$ & $z=0$ \\
\tableline
SCDM  & $-2.25 \pm 0.04$  & $-2.18 \pm 0.03$  & $-1.98 \pm 0.02$ & $-2.02 \pm 0.09$ & $-2.12 \pm 0.05$ \\
CCDM  & $-1.98 \pm 0.02$  & $-1.97 \pm 0.02$ & $-1.97 \pm 0.02$ &     -    &     -   \\
OCDM  & $-2.18 \pm 0.04$  & $-2.09 \pm 0.03$ & $-2.00 \pm 0.02$ &     -    &     -   \\
LCDM  & $-2.21 \pm 0.08$  & $-2.28 \pm 0.04$ & $-2.06 \pm 0.03$ &     -    &     -   \\
TCDM  & $-2.82 \pm 0.19$  & $-2.46 \pm 0.14$ & $-2.10 \pm 0.07$ &     -    &     -   \\

\end{tabular}
\end{center}
\end{table}

\begin{figure}
\centerline{
\epsfxsize=4.5truein
\epsfbox[95 415 465 735]{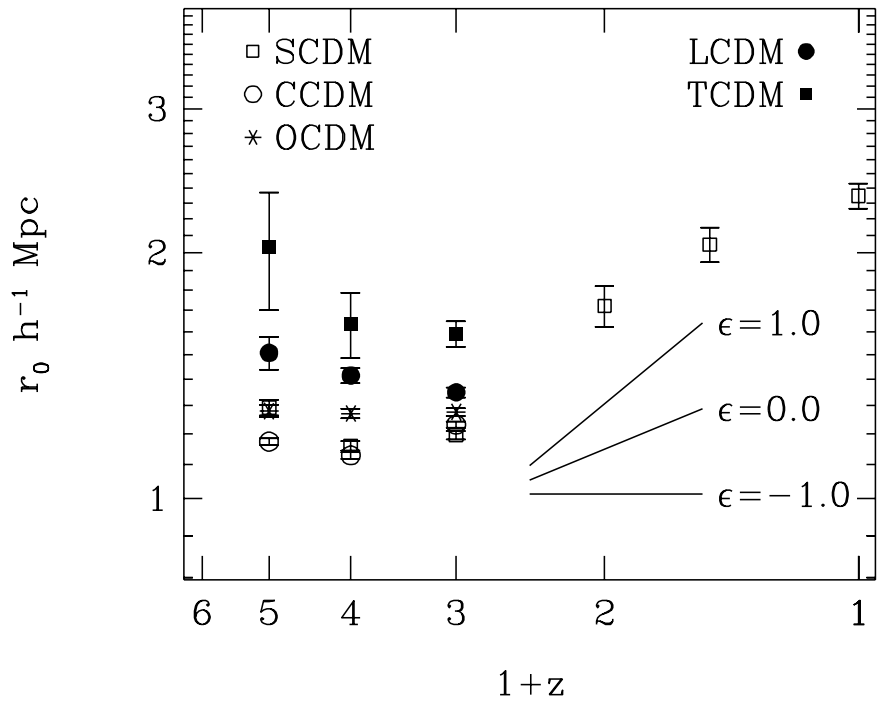}
}
\caption{              
\label{fig:growth}
Redshift evolution of the galaxy correlation length.
Solid lines show the expected growth rates for different values of $\epsilon$ 
(eq. [\ref{eqn:growth}]).
}
\end{figure}

Figure~\ref{fig:growth} plots the redshift evolution of the
correlation lengths $r_0$.  Within a given model, there is little
change in $r_0$ from $z=4$ to $z=2$.  To the extent that there is
evolution over this redshift interval, $r_0$ declines slightly in 
the models with the weakest mass fluctuations, TCDM and LCDM,
and stays flat or rises slightly in the models with stronger mass
fluctuations, SCDM, OCDM, and CCDM.  The SCDM points at
$z=1$, 0.5, and 0 are taken from the lower resolution simulation
described in \S 2.  At low redshift, the correlation length
does increase with time, as gravitational clustering pulls galaxies
and clusters into denser and larger structures.

In observational analyses and theoretical discussions, the
evolution of the correlation function is often parameterized
as a power law in redshift, 
\begin{equation}
\xi(r,z) = (r/r_0)^{\gamma}(1 + z)^{-(3+\epsilon+\gamma)}
\label{eqn:growth}
\end{equation}
where $r_0$ is the comoving correlation length (e.g.,
\cite{gp77}; \cite{ebktg91}).
This equation in turn implies 
\begin{equation}
r_0(z) =  r_0 (1 + z)^{1 + (3 + \epsilon)/\gamma}.
\label{eqn:r0growth}
\end{equation}
Two interesting, though not necessarily limiting, cases are a
clustering pattern fixed in comoving coordinates, implying
$\epsilon = -3 - \gamma$, and a clustering pattern that is
dynamically bound and stable in physical coordinates, implying
$\epsilon = 0$ (\cite{gp77}; note that this is not the same as
a fixed physical correlation length because the contrast 
of structure grows as the background density drops.)
Linear fluctuation growth in an Einstein-de Sitter universe
corresponds to $\epsilon = -1-\gamma$.

The solid lines in Figure~\ref{fig:growth} show the $r_0$ evolution
for $\epsilon=-1$, 0, and $+1$, corresponding to fixed comoving
clustering, fixed physical clustering, and linear growth, respectively,
for $\gamma=-2$.  At high redshifts, $r_0$ is nearly fixed in
comoving coordinates, with $\epsilon=-1$.  For the SCDM model
at lower redshift, we find evolution in between fixed physical
clustering and linear growth.  However, it is clear from 
Figure~\ref{fig:growth} that the power law parameterization
of equation~(\ref{eqn:growth}) can at best be an approximation
over a rather narrow range of redshifts.  Between $z=4$ and
$z=0$, the predicted evolution of the galaxy correlation 
length is not described by a single power law, and in some
cases it is not even monotonic.

\subsection{Bias Factors}
\bigskip

It is obvious from Figures~\ref{fig:circle} and~\ref{fig:xi1}
that the clustering of the SPH galaxies is biased relative to 
the clustering of the dark matter.  In order to quantify this
bias with a statistically robust measure, we define a bias factor
$b(R)$ from the ratio of $J_3$ integrals:
\begin{equation}
b(R) = \left(J_{\rm 3,gal}(R) \over J_{\rm 3,dark}(R)\right)^{0.5} ~,
\quad{\rm where}
\label{eqn:b}
\end{equation}
\begin{equation}
J_3(R) = \int_0^R \xi(r) r^2 dr
\label{eqn:j3}
\end{equation}
(Peebles 1980).
The quantity $4\pi\nbar J_3(R)$ is the mean number of galaxies
in excess of random within a distance $R$ of a randomly chosen
galaxy (where $\nbar$ is the mean galaxy number density).
We do not examine the scale dependence of bias in this
paper but instead define a single bias factor $b$ from 
equation~(\ref{eqn:b}) with $R=4\hmpc$.  We compute $J_3$ in the
simulations by simply counting the average number of pairs
out to $R=4\hmpc$, subtracting the Poisson contribution to 
obtain the number of excess (correlated) pairs, and dividing
by $4\pi \nbar$.  For a power law correlation function with $\gamma=-2$,
the bias factor $b$ is equal to the ratio $r_{0,{\rm gal}}/r_{0,{\rm dark}}$
of the galaxy and dark matter correlation lengths, and $J_3(R) = r_0^2 R$.

\begin{figure}
\centerline{
\epsfxsize=4.5truein
\epsfbox[95 415 465 735]{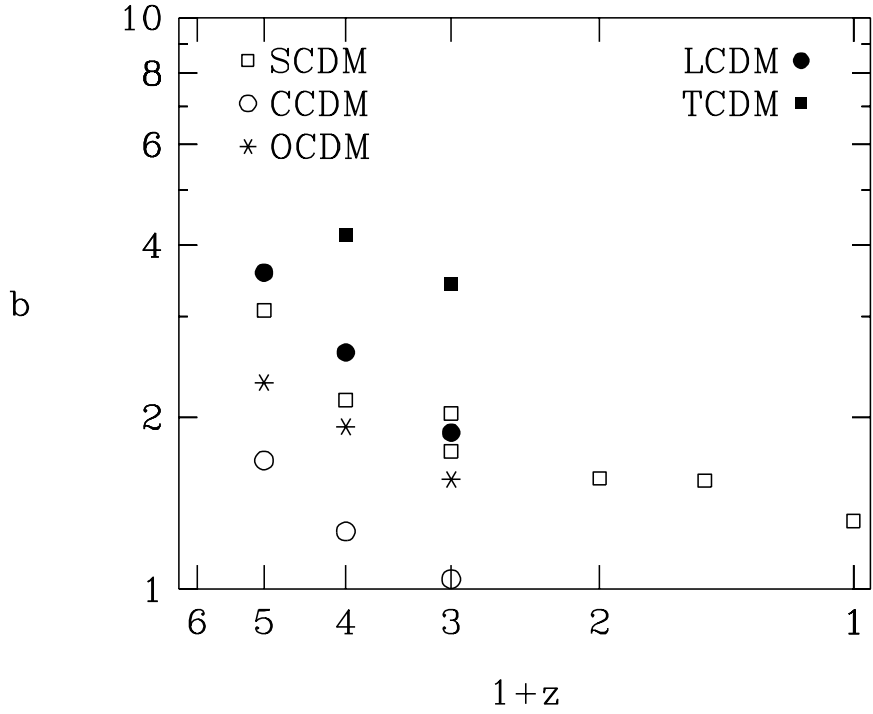}
}
\caption{              
\label{fig:bz}
Redshift evolution of the bias factor, $b$.
The value of $b$ is computed from eq.~(\ref{eqn:b}) using all SPH galaxies.
Results for SCDM at $z=0$, 0.5, 1, and the upper SCDM point at $z=2$,
come from the lower resolution
SCDM simulation.  The TCDM model is not plotted at $z=4$ because
its correlation function is too noisy.
}
\end{figure}

Figure~\ref{fig:bz} shows the redshift evolution of the bias factor
in all of the models.  In every case the bias factor drops between
$z=4$ and $z=2$, and it continues to drop from $z=2$ to $z=0$ in
the SCDM model.  This is the behavior expected from Figure~\ref{fig:xi1},
which shows that the galaxy correlation function remains roughly fixed while
the dark matter correlation function grows in time, steadily
reducing the bias between them.  At $z=3$ the bias factor ranges from
$b=1.3$ in CCDM to $b=4.2$ in TCDM.  

\begin{figure}
\centerline{
\epsfxsize=4.5truein
\epsfbox[65 0 550 740]{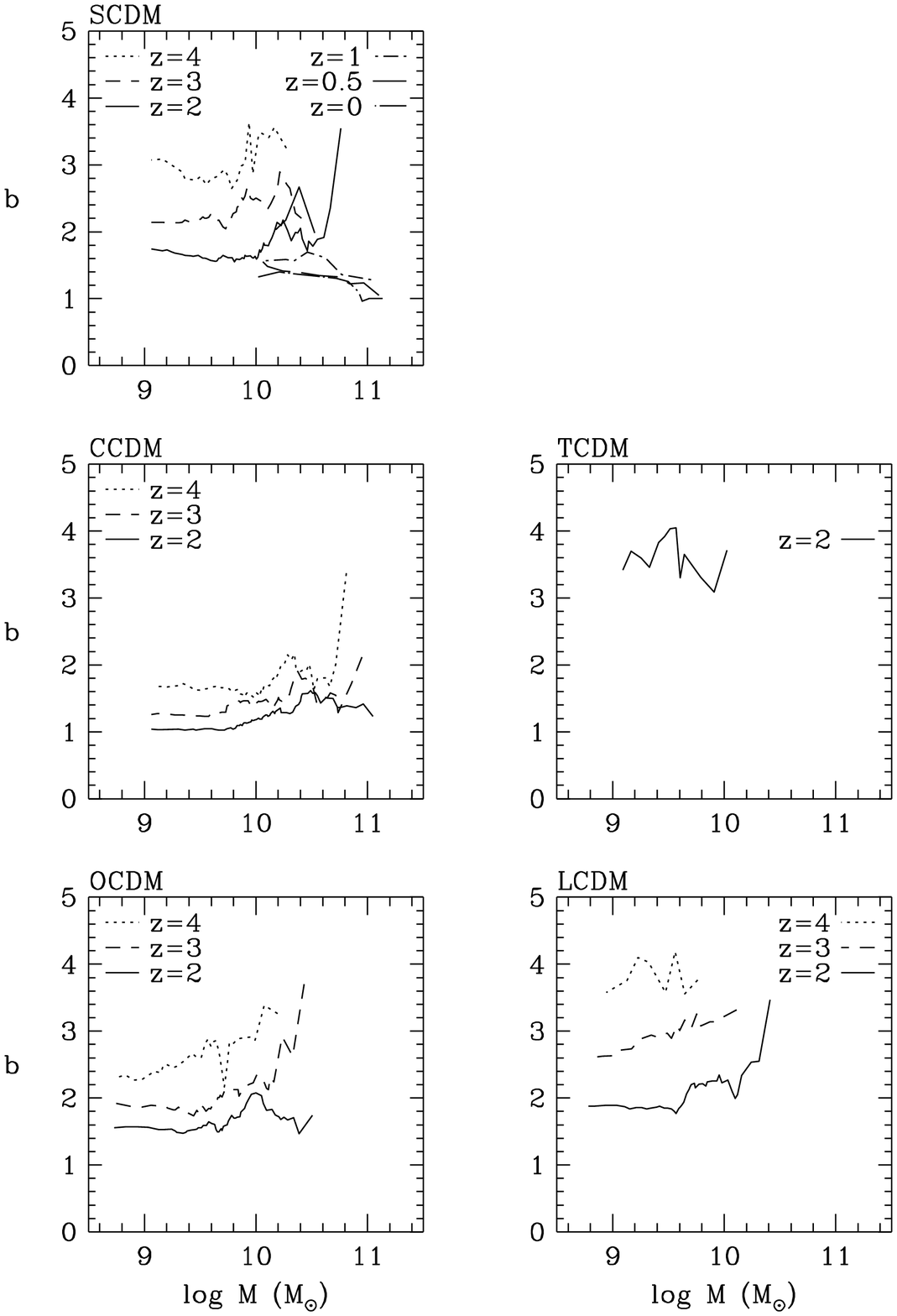}
}
\caption{              
\label{fig:bm}
Mass dependence of the bias factor.  The bias factor $b(M)$ for all
galaxies with baryonic mass greater than $M$ is plotted as a function of $M$,
at redshifts $z=4$, 3, and 2.  Results from the lower resolution SCDM
simulation are shown at $z=2$, 1, 0.5, and 0.
}
\end{figure}

In the one-halo-one-galaxy model, the amplitude of bias is expected
to depend on the mass of the parent halos (\cite{mow96}).  We might
reasonably expect this dependence to carry over into a dependence
of bias on galaxy baryon mass.  In Figure~\ref{fig:bm}, therefore,
we plot the bias factor as a function of baryon mass: $b(M)$ is
the bias factor computed from the $J_3$ integral of all galaxies
whose baryon mass exceeds $M$.  We plot $b$ values only if the
number of correlated pairs (pairs in excess of the Poisson prediction)
exceeds 6.25, so that the shot noise error in the computation of
$b$ is less than 25\%.  The low mass cutoff of the curves corresponds
to our mass resolution limit, which is lower in the LCDM and OCDM
models because the lower $\Omega_b$ implies a lower SPH particle mass.
Prior to $z=2$, the results for the TCDM model are too noisy
to allow useful examination of the dependence of bias on galaxy mass.

\begin{figure}
\centerline{
\epsfxsize=4.5truein
\epsfbox[65 0 550 740]{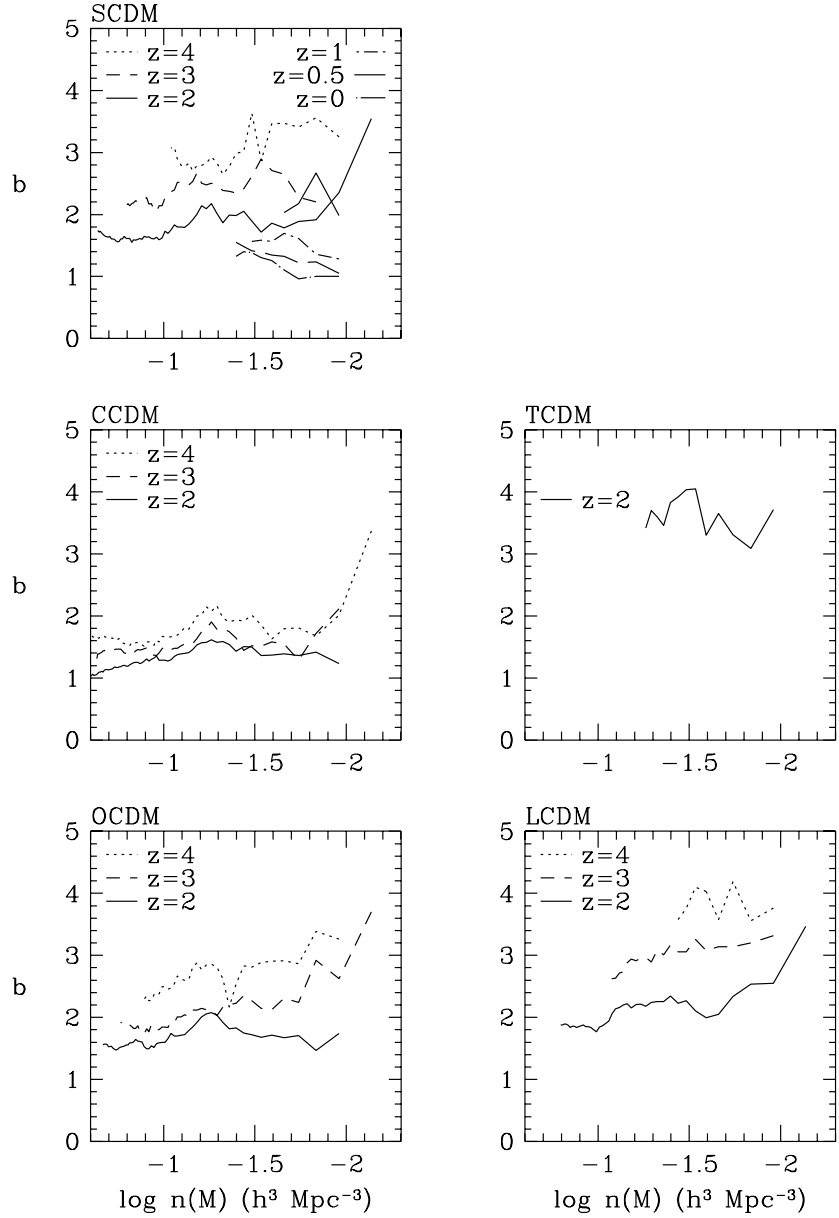}
}
\caption{              
\label{fig:bn}
Same as Fig.~\ref{fig:bm}, except that $b(M)$ is plotted against $n(M)$,
the comoving number density of galaxies with baryonic mass greater than $M$.
}
\end{figure}

In most cases, Figure~\ref{fig:bm} shows a trend of increasing $b$
with increasing galaxy mass.  The trend is usually weak at low
masses, but in several cases we find considerably stronger bias
for the most massive galaxies in the simulation.  While mass is
likely to be the physical parameter most closely related to $b$, 
it is difficult to infer the mass of a Lyman-break galaxy directly
from observations.  {\it If} the star formation rate (and hence
the rest frame UV luminosity) is tightly correlated with the galaxy
baryon mass, then the Lyman break galaxies that make it into a
magnitude limited sample will be the most massive galaxies present
at the observed redshift.  With this assumption, one can use space
density as an observable proxy for galaxy mass.  In Figure~\ref{fig:bn},
we plot $b(M)$ as a function of $n(M)$, the comoving space
density of galaxies with baryonic mass greater than $M$.
Unfortunately, our box is not large enough to allow statistically
reliable measures of $b$ at the space density of the ASGDPK sample.
The problem will eventually be rectified with larger volume simulations,
and with observational studies that probe clustering further down the
luminosity function, but for now the gap between the space densities
of the simulated and observed samples is an uncertainty we must live with.

The predicted bias of high redshift galaxies shown in 
Figures~\ref{fig:bz}--\ref{fig:bn} is generally (but not always)
higher than the estimated bias factor of galaxies today.
The determination of $b$ at $z=0$ is intimately connected to
the determination of $\Omega$, and virtually all current analyses
measure a combination of $b$ and $\Omega$ rather than $b$ alone.
For example, the masses of observed galaxy clusters imply
that the rms mass fluctuation in $8\hmpc$ spheres is 
$\sigma_8 \approx 0.5 \Omega^{-0.6}$
(e.g., \cite{white93}; \cite{eke96}), which in 
combination with the measured fluctuation amplitude $\sigma_{8,I}=0.7$
for IRAS-selected galaxies (\cite{fisher94}) implies a bias factor
$b_I \approx 1.4\Omega^{0.6}$.  The VELMOD peculiar velocity analysis
of Willick \& Strauss (1998) and redshift-space distortion analysis
of Cole, Fisher, \& Weinberg (1995) imply a larger bias factor
$b_I \approx 2\Omega^{0.6}$, while the POTENT peculiar velocity
analysis of Sigad et al.\ (1998) implies a lower value, 
$b_I \approx 1.12 \Omega^{0.6}$ (see Strauss \& Willick 1995 for
a summary of numerous other estimates of $b_I$).
The {\it relative} bias of 
optical galaxies to IRAS galaxies can be estimated from direct
comparisons of their correlation functions, $b_O/b_I \approx 1.4$
at the $8\hmpc$ scale (\cite{strauss92}; \cite{fisher94};
\cite{moore94}; \cite{peacock96}), though there are also
strong indications that the clustering of optical galaxies
depends on luminosity (\cite{hamilton88}; \cite{park94}; \cite{loveday95}).
If we accept the mounting but still circumstantial evidence for
a low density universe and therefore adopt $\Omega \approx 0.4$,
then a reasonable reading of the current constraints
would be $b_I \approx 0.7$ and $b_O \approx 1$ at $z=0$.

\subsection{Finite Volume Corrections}
\bigskip

As we have already mentioned, the absence of fluctuations on 
scales larger than the fundamental mode of our $11.111\hmpc$ 
simulation cube makes the galaxy correlation length in our
simulations artificially low and the correlation function 
artificially steep.  We can estimate the size of this effect
by using purely gravitational N-body simulations with larger
volume and lower spatial resolution.  Specifically, for each cosmological
model we perform six simulations using the particle-mesh (PM) code of 
Park (1990; see also \cite{park91}).  Each simulation uses $192^3$
particles and a $192^3$ density-potential mesh to compute the
gravitational clustering in a cube $66.66\hmpc$ on a side, a volume
216 times larger than that of our SPH simulations.
The spatial resolution 
(size of a mesh cell) in the PM simulations is $347(1+z)^{-1}\hkpc$
(physical units), compared to the $3(1+z)^{-1}\hkpc$ gravitational
force softening used in the TreeSPH simulations.  
We measure the dark matter correlation lengths in the PM simulations
using least squares power law fits as before, but we restrict
the fit to spatial scales larger than twice the spatial resolution
of the PM simulations.  We have carried out PM simulations of
$11.111\hmpc$ cubes with the same resolution (i.e., $32^3$ particles on
a $32^3$ mesh) to confirm that the PM simulations reproduce the
dark matter correlation function of the TreeSPH simulations on
these scales, and to confirm that the mass correlation function obtained
for the specific set of initial conditions used in the TreeSPH
simulations is similar to the mean mass correlation function obtained
in volumes of this size.

We use the results of the larger volume simulations to compute
a ``finite volume $r_0$ correction factor'' for each model at each redshift:
the ratio of the dark matter correlation length in the large volume simulations
to the dark matter correlation length in the corresponding TreeSPH simulation.
These correction factors are listed in Table~\ref{tbl-4}; they
range from 1.4 to 2.3, with most values in the range 1.9 to 2.2.
Thus, our finite simulation volume causes us to underestimate the
dark matter correlation length by about a factor of two.
The slope of the correlation functions in the larger volume simulations
is also shallower, typically $\gamma \approx 1.8$ instead of
$\gamma \approx 2.0$. 

\begin{table}
\caption{Finite volume $r_0$ correction factors} \label{tbl-4}
\begin{center}\scriptsize
\begin{tabular}{crrrrrrrrrrr}
Model & $z=4$ & $z=3$ & $z=2$ & $z=1$ & $z=0$ \\
\tableline
SCDM  & $1.59 \pm 0.02$  & $1.97 \pm 0.01$  & $1.90 \pm 0.01$ & $1.98 \pm 0.007$ & $2.0
8 \pm 0.004$ \\
CCDM  & $1.87 \pm 0.01$  & $1.89 \pm 0.01$ & $1.90 \pm 0.01$ &     -    &     -   \\
OCDM  & $2.06 \pm 0.02$  & $2.11 \pm 0.01$ & $2.17 \pm 0.01$ &     -    &     -   \\
LCDM  & $2.09 \pm 0.02$  & $2.20 \pm 0.02$ & $2.28 \pm 0.01$ &     -    &     -   \\
TCDM  & $1.38 \pm 0.03$  & $1.54 \pm 0.03$ & $2.11 \pm 0.02$ &     -    &     -   \\

\end{tabular}
\end{center}
\end{table}

\begin{figure}
\centerline{
\epsfxsize=4.5truein
\epsfbox[95 415 465 735]{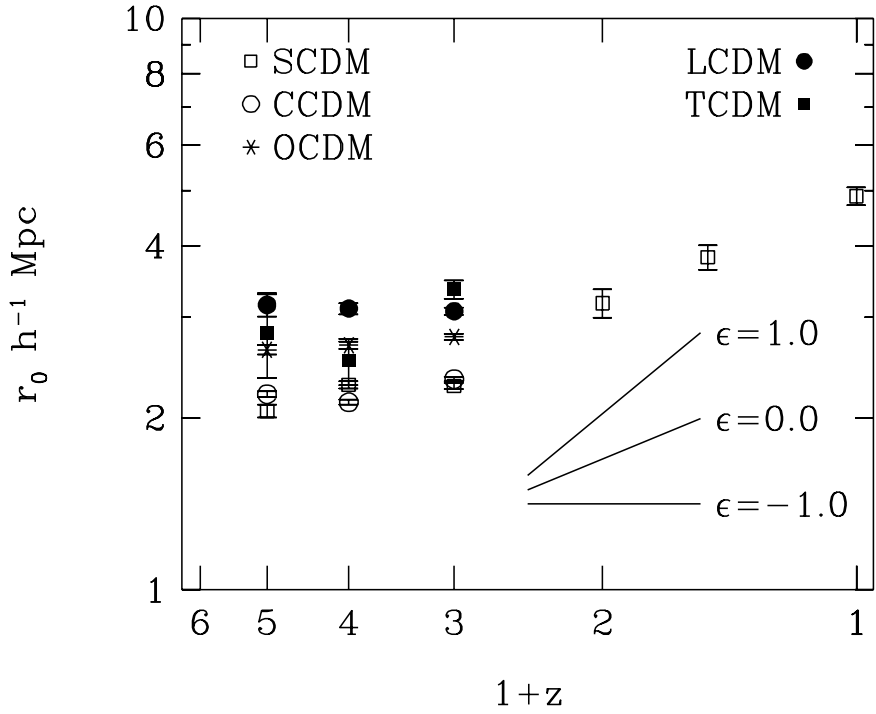}
}
\caption{              
\label{fig:growth2}
Redshift evolution of the galaxy correlation length,
as in Fig.~\ref{fig:growth}, with correlation lengths
corrected for the effects of the finite simulation volume.
Solid lines show the expected growth rates for different values of $\epsilon$ 
(eq. [\ref{eqn:growth}]).
}
\end{figure}

In order to estimate the {\it galaxy} correlation lengths that we 
would obtain in larger volume simulations, we assume that our
$11.111\hmpc$ simulations correctly calculate the {\it bias}
between galaxies and dark matter.  While we do not expect this
assumption to be precisely correct, since, for example, the abundance
of peaks in the initial conditions is modulated by large scale
waves (\cite{kaiser84}; \cite{bbks86}), it is likely to be
a reasonable first approximation.  With this assumption, we simply
multiply the galaxy correlation lengths from Table~\ref{tbl-2} by
the correction factors in Table~\ref{tbl-4} to obtain the corrected
galaxy correlation lengths, which are listed in Table~\ref{tbl-5}.
At $z \geq 2$, the corrected correlation lengths range from $2.1\hmpc$
to $3.4\hmpc$.  Figure~\ref{fig:growth2} is a repeat of 
Figure~\ref{fig:growth}, but it uses the corrected correlation lengths.
Since the correction factors are not strongly dependent on redshift,
the evolutionary trends are the same as those found previously.
The $r_0$ values listed in Table~\ref{tbl-5} and plotted in
Figure~\ref{fig:growth2} are
our best estimates of the predicted correlation lengths for 
Lyman-break galaxies with $v_c \geq 100\;\kms$ ($\Omega=1$)
or $v_c \geq 70\;\kms$ ($\Omega=0.4$).
Because the strength of the galaxy clustering is an increasing function of 
galaxy mass (Figure~\ref{fig:bm}), we expect the correlation
lengths of galaxies in existing Lyman-break samples to be
somewhat larger, a point that we will return to shortly.

\begin{table}
\caption{Finite volume corrected correlation lengths, $r_0$ in $\hmpc$ } \label{tbl-5}
\begin{center}\scriptsize
\begin{tabular}{crrrrrrrrrrr}
Model & $z=4$ & $z=3$ & $z=2$ & $z=1$ & $z=0$ \\
\tableline
SCDM  & $2.06 \pm 0.05$  & $2.28 \pm 0.04$  & $2.27 \pm 0.03$ & $3.17 \pm 0.18$ & $4.89
 \pm 0.18$ \\
CCDM  & $2.20 \pm 0.03$  & $2.13 \pm 0.02$ & $2.34 \pm 0.02$ &     -    &     -   \\
OCDM  & $2.63 \pm 0.05$  & $2.68 \pm 0.04$ & $2.77 \pm 0.03$ &     -    &     -   \\
LCDM  & $3.15 \pm 0.15$  & $3.11 \pm 0.07$ & $3.07 \pm 0.05$ &     -    &     -   \\
TCDM  & $2.82 \pm 0.47$  & $2.52 \pm 0.24$ & $3.36 \pm 0.13$ &     -    &     -   \\

\end{tabular}
\end{center}
\end{table}

\section{Discussion}
\bigskip

Our basic results are easy to summarize.  While the five CDM models that we 
examine in this paper differ in their cosmological parameters, differ
(substantially) in their mass clustering amplitudes, and differ (slightly)
in their initial power spectrum shapes, they predict remarkably similar
galaxy clustering at $z=4$, 3, and 2.  The galaxy correlation functions
show almost no evolution over this redshift range, even though the mass
correlation functions grow steadily.  In most cases the high redshift 
galaxies are strongly biased tracers of the underlying mass, but the bias
factor evolves with redshift and varies from model to model, roughly in 
inverse proportion to the amplitude of mass fluctuations.  The more massive
galaxies tend to be more highly clustered, though the trend is fairly
weak over the mass range for which our $\xi(r)$ measurements are statistically
robust.  In our lower resolution SCDM simulation evolved to low redshift,
the strength of galaxy clustering grows from $z=2$ to $z=0$.

The evolution of the galaxy correlation function in our simulations is 
qualitatively similar to that in analytic models that identify galaxies
with high peaks of the initial density field (\cite{bbks86}; \cite{bagla98b}).
The galaxies are ``born'' strongly clustered and highly biased, and at early
times the dark matter simply falls into the structures where the galaxies
already reside, driving down the bias factor without substantially changing
the galaxy correlation function.  Eventually the gravitational clustering
of the dark matter begins to change the galaxy distribution, and the
galaxy $\xi(r)$ increases, as seen at low redshift in the SCDM simulation.
The evolution of $\xi(r)$ is similar to that predicted in Fry's (1996)
analytic model, in which an initially biased galaxy distribution follows
the dark matter velocity field.  Galaxy mergers and ongoing galaxy formation
complicate the picture, however, as shown (for example) by the fact that
the dark matter clustering actually overtakes the galaxy clustering at
$z=2$ in our CCDM simulation.  Tegmark \& Peebles (1998) have recently
described an elegant generalization of Fry's (1996) approach that can 
incorporate ongoing galaxy formation and stochastic bias, but this is presently 
a framework for calculation rather than a predictive model of clustering
evolution.  We suspect that the semi-analytic/N-body approach
(\cite{governato98}; \cite{kauffmann98}) would yield similar evolution
predictions, but there are no published evolution results to which
we can compare at present.

Our prediction of a strongly biased galaxy population at high redshift
is similar to the prediction of models that identify Lyman-break galaxies
directly with the high mass tail of the dark halo population
(\cite{mof96}; \cite{bagla98a}b; \cite{coles98}; \cite{jing98};
\cite{wechsler98}).  Like the halo-based calculations, the SPH simulations
predict a trend of increasing bias with increasing galaxy mass.
The trend appears to be somewhat weaker in the SPH simulations than in
the analytic halo model of Mo \& White (1996); in particular, we almost
always find a positive galaxy bias even with a low circular velocity
threshold and a correspondingly high galaxy space density.  We also find
a positive bias for friends-of-friends halos in an N-body simulation started
from the same initial conditions as our SCDM SPH simulation
(Figures~\ref{fig:halo} and~\ref{fig:xihalo}).  The halos show a slightly
stronger trend of bias with space density than that of the SPH galaxies,
at least for the commonly used linking parameter $d_l=0.2$ (the halo
clustering results also depend on the algorithm and parameters used for
halo identification).  It is not clear that our findings for halos conflict
with previous results, since none that we know of show $\xi(r)$ for 
low mass, high redshift halos in the non-linear regime, where $\xi(r) \ga 1$.  
We also note that Jing (1998) finds that in the quasi-linear regime
the bias of low mass N-body halos is stronger than the Mo \& White (1996)
analytic model predicts, though he still finds that the low mass halos
are antibiased and finds agreement with the Mo \& White (1996) formula
for high mass halos.  For now, the most we can say is that the analytic
formalism for the bias of halos in the linear regime should be used with
some caution, especially if one is interpreting the clustering of galaxies
in the non-linear regime, where the Lyman-break galaxy correlation function
is presently measured.

Our best estimates of the predicted galaxy correlation lengths are those
listed in Table~\ref{tbl-5} and plotted in Figure~\ref{fig:growth2}.
At $z=3$, they lie in the range $2.1-3.1\hmpc$.  These predictions apply
to the full population of galaxies above our resolution limit.  The more
massive galaxies are more strongly clustered.  One can find the predicted 
correlation length for galaxies with mass greater than $M$ by multiplying
the value in Table~\ref{tbl-5} by the ratio 
$[b(M)/b({\rm all})]^{2/|\gamma|} \approx b(M)/b({\rm all})$, read
from Figure~\ref{fig:bm}, where $b({\rm all})$ is the bias factor for
the full galaxy population (the left end of the $b(M)$ curve).
The increase in $r_0$ is often $30-50$\% or even more for the most massive
galaxies, but it is rather uncertain because of our limited statistics.

{}From an analysis of the angular correlation function of Lyman-break
galaxy candidates, Giavalisco et al.\ (1998a) find $\gamma = -2.0 \pm 0.3$
and a correlation amplitude that corresponds to $r_0=2.1 \pm 0.5\hmpc$
for $\Omega=1$ or $2.8 \pm 0.7\hmpc$ for $\Omega=0.4$ (with either
$\Lambda=0$ or $\Lambda=0.6$).  ASGDPK estimate a significantly higher
correlation length, $r_0=4 \pm 1\hmpc$ ($\Omega=1$),
$r_0=5 \pm 1\hmpc$ ($\Omega=0.2$, $\Lambda=0$), or
$r_0=6 \pm 1\hmpc$ ($\Omega=0.3$, $\Lambda=0.7$), from a counts-in-cells
analysis of spectroscopically confirmed Lyman-break 
objects.\footnote{
Scaling the $\Omega=1$ results by the ratios of angle-distance
or redshift-distance relations yields $r_0 \approx 5.5 \pm 1.4\hmpc$ for the
parameters of our OCDM and LCDM models; the scaling in ASGDPK is
somewhat different because they estimate $r_0$ separately for each
cosmology using cubical cells whose redshift depth matches the angular
size of the survey field.}
The effective redshift of both determinations is $z \approx 3$, and
the difference between them provides an indication of the current
level of measurement uncertainties.  

Quantitative comparison between our results and these data is hampered
by the non-overlapping space densities of the theoretical and observational
samples.  However, it at least seems clear that the models considered
in this paper have no difficulty explaining the strong observed clustering
of Lyman-break galaxies.  Indeed, with a reasonable extrapolation of the
$b(M)$ vs. $n(M)$ curves in Figure~\ref{fig:bn}, it seems that some of
the models might overpredict even the stronger clustering found by ASGDPK.
Interpreting their results in light of the Mo \& White (1996) halo
biasing formula, ASGDPK conclude that their data favor a $\Gamma=0.2$
shape parameter for the primordial mass power spectrum over $\Gamma=0.5$,
but we find that the predicted galaxy correlation function in the SPH
simulations (admittedly at a space density higher than that of the
ASGDPK sample) depends only weakly on the shape or amplitude of the 
underlying mass power spectrum.  

Steidel et al.\ (1998b) report preliminary results 
(from Giavalisco et al.\ 1998b) for the clustering of lower luminosity
Lyman-break galaxies, derived from the angular correlation function
of $U$-band dropouts in the Hubble Deep Field (HDF; \cite{williams96}).
The estimated space density of this sample is 45 times higher than that
of the ASGDPK sample, and the estimated bias factor is 3 times lower
(\cite{steidel98}, figure 7).  Since the correlation length scales
as $r_0 \propto b^{2/|\gamma|}$, this lower bias factor implies
(in combination with the ASGDPK results) $r_0 \sim 1.2\;\hmpc$
for $\Omega=1$ and $r_0 \sim 1.6\;\hmpc$ for $\Omega=0.4$,
assuming $\gamma=-1.8$.  At the space density of the HDF sample there would
be $\sim 120$ galaxies in our $\Omega=1$ simulation boxes and $\sim 50$
galaxies in our $\Omega=0.4$ simulation boxes, so it is reasonable to
compare these numbers directly to the correlation lengths in Table~\ref{tbl-5}.
The simulations appear to predict excessively strong clustering
for the Giavalisco et al.\ (1998b) HDF sample, particularly with $\Omega=1$.

Given the current observational and theoretical uncertainties, it is
premature to make much of this conflict, which at present has only 
marginal statistical significance.  One source of theoretical
uncertainty, the small size of our simulation volumes, is purely technical,
and it will be remedied in the near future with larger simulations.
The other source of theoretical uncertainty, the unknown relation 
between galaxy baryon mass and rest-frame UV luminosity, is more
fundamental.  Because the clustering strength depends on galaxy
mass (Figure~\ref{fig:bm}), the correlation function predicted for
a given Lyman-break sample depends on how that sample probes the
galaxy mass function.  Our conversion from galaxy mass to space density
in Figure~\ref{fig:bn} implicitly assumes that the mass-luminosity 
relation is monotonic, so that a magnitude limited sample selects
galaxies at the top end of the mass function.
If episodic star formation or stochastic dust extinction introduce
large scatter into the relation between galaxy mass and UV luminosity,
then the predicted relation between clustering strength and space
density will be flatter and the predicted clustering for bright Lyman-break
galaxy samples will be lower.

Some papers on high redshift structure (e.g., \cite{coles98}; \cite{jing98}) 
conclude
that measurements of Lyman-break galaxy clustering will in time provide
valuable tests of cosmological models.  Other authors have argued
that these measurements will ultimately teach us more about the nature
of Lyman-break galaxies themselves (e.g., SADGPK; Giavalisco et al.\ 1998a;
ASGDPK).  In this debate, we side firmly with the latter group.
We have examined models with very different mass clustering properties,
but the effects of biased galaxy formation conspire to erase the
differences between them.  This ``conspiracy'' is not as suspicious
as it initially seems; it simply reflects the fact that galaxies form
at similar ``special'' locations in the density fields of the
different models and therefore have similar spatial correlations
(Figure~\ref{fig:circle}).  The Ly$\alpha$ forest provides a 
more straightforward route to testing the high redshift structure
predictions of cosmological theories because the physics of the
absorbing medium is relatively simple and leads to a direct
relation between an observable quantity, the Ly$\alpha$ optical depth,
and the underlying mass density (WKH).  Indeed, the determination of
the high redshift mass power spectrum by the method of Croft et al.\ (1998)
will yield the bias factor of Lyman-break galaxies as a by-product.

Given the complications introduced by bias, we think it is
better to regard Lyman-break galaxy clustering as a probe of galaxy
formation physics rather than a test of cosmological models.
The fact that the SPH simulations (and other approaches based on
peaks, halos, or semi-analytic models) account naturally for the observed
strong clustering is already an encouraging success for the general
scenario of hierarchical galaxy formation from Gaussian primordial
density fluctuations.  Comparison between the predicted trend of
clustering strength against galaxy mass and the observable trend of
clustering strength against luminosity can provide a more detailed
test of theoretical models and an indication of whether the 
instantaneous star formation rate is tightly correlated with galaxy mass.
This line of argument has already been pursued by SADGPK and ASGDPK
in the context of halo models.  In the context of SPH simulations,
the trend of clustering with UV luminosity can ultimately help to
test whether our adopted paradigm for galactic scale star formation
is basically correct or whether galaxy interactions or stochastic
bursts have a strong influence on a high redshift galaxy's 
star formation rate.  The study of high redshift galaxy clustering
is still in its early phases, but progress in both observation
and theory has been remarkably rapid, and it looks likely to continue.

\acknowledgments 

We thank Eric Linder, Max Pettini, and Chuck Steidel 
for useful discussions and Houjun Mo 
and Yipeng Jing for helpful exchanges about halo clustering.
We thank Changbom Park for allowing us to use his PM N-body code.
This work was supported
by NASA Astrophysical Theory Grants NAG5-3922, NAG5-3820, and NAG5-3111,
by NASA Long-Term Space Astrophysics Grant NAG5-3525, and by the NSF under
grants ASC93-18185 and ACI96-19019.  The simulations were performed at
the San Diego Supercomputer Center.

\vfill\eject

\end{document}